\documentclass[useAMS,usenatbib,onecolumn]{mn2e}
\usepackage{graphicx,amssymb,natbib}

\def\fexxiv{Fe~{\sc xxiv}}
\def\fexxv{Fe~{\sc xxv}}

\def\fexxi{Fe~{\sc xxi}}

\title[Solar Fe abundance]
  {The Solar Photospheric-to-Coronal Fe abundance from X-ray Fluorescence Lines}
\author[K. J. H. Phillips]
  {K. J. H. Phillips$^1$ \\
  $^1$Mullard Space Science Laboratory, University College London, Holmbury St Mary, Dorking RH6 5NT, UK}

\date{Submitted 2011 November}

\pagerange{\pageref{firstpage}--\pageref{lastpage}} \pubyear{2011}

\begin{document}

\label{firstpage}

\maketitle
\input epsf

\begin{abstract}
The ratio of the Fe abundance in the photosphere to that in coronal flare plasmas is determined by X-ray lines within the complex at 6.7~keV (1.9~\AA) emitted during flares. The line complex includes the He-like Fe (\fexxv) resonance line $w$ (6.70~keV) and Fe K$\alpha$ lines (6.39, 6.40~keV), the latter being primarily formed by the fluorescence of photospheric material by X-rays from the hot flare plasma. The ratio of the Fe K$\alpha$ lines to the \fexxv\ $w$ depends on the ratio of the photospheric-to-flare Fe abundance, heliocentric angle $\theta$ of the flare, and the temperature $T_e$ of the flaring plasma. Using high-resolution spectra from X-ray spectrometers on the {\em P78-1} and {\em Solar Maximum Mission} spacecraft, the Fe abundance in flares is estimated to be $1.6\pm 0.5$ and $2.0 \pm 0.3$ times the photospheric Fe abundance, the {\em P78-1} value being preferred as it is more directly determined. This enhancement is consistent with results from X-ray spectra from the {\em RHESSI} spacecraft,  but is significantly less than a factor 4 as in previous work.
\end{abstract}


% Section 1: Introduction
\section{Introduction}

The solar iron abundance remains a subject for discussion in the literature, both the value obtained from photospheric spectra and that from coronal spectra, generally in the X-ray or extreme ultraviolet spectral regions. Recent photospheric abundance determinations \citep{asp09} using three-dimensional photospheric models give $A({\rm Fe}) = 7.50 \pm 0.04$ \citep{asp09} and $7.52 \pm 0.06$ \citep{caf11} (abundances are here expressed logarithmically, on a scale where $A({\rm H}) = 12)$, less by up to 60\% (0.2~dex) than values current as recently as the mid-1990s \citep{bla95}. Since the 1980s, it has become widely recognized that solar coronal element abundances differ from photospheric by factors that depend on the first ionization potential (FIP) of the element, with the coronal abundances of low-FIP (i.e. $\lesssim 10$~eV) being enhanced but with the coronal abundances of high-FIP elements that are similar to their photospheric abundances. The reviews by \cite{fel92} and \cite{fel00} give enhancement factors for some low-FIP elements of up to 4. However, low-altitude coronal flare plasmas and other energetic events are stated to have photospheric abundances, while \cite{fel90} find that the corona directly above a sunspot also has photospheric abundances but neighbouring coronal regions have low-FIP elements with enhanced abundances. Iron, with a FIP of 7.9~eV, is expected to have a low-FIP element behaviour. A recent analysis \citep{phi11} of solar flare spectra observed by the {\em Reuven Ramaty High Energy Solar Spectroscopic Imager} ({\em RHESSI}) gives a rather different picture. {\em RHESSI} spectra cover an energy range $\sim 3$~keV to 17~MeV, and have a spectral resolution of $\Delta E \sim 1$~keV; in the soft X-ray range ($\sim 3-10$~keV, $1.2-4$~\AA), a strong complex of Fe lines, forming an emission line feature at 6.7~keV ($1.85$~\AA), is evident on a continuum made up of free--free and free--bound emission. Under certain circumstances -- late in the flare development, with the {\em RHESSI} thin attenuators in place -- an isothermal assumption for the flare's emitting plasma appears to be valid, and with electron temperature $T_e$ and emission measure ($EM= N_e^2 V$, where $N_e = $ electron density, $V = $ the emitting volume) estimated from the continuum, the Fe abundance can be determined. Analysis of nearly 2000 spectra observed during 20 flares led to the value $A({\rm Fe}) = 7.91 \pm 0.10$, with very little variation from flare to flare. This abundance is a factor of $2.6 \pm 0.6$ more than recent photospheric abundance estimates, significantly less than 4.

\cite{phi94} describe a means of obtaining the photospheric-to-flare Fe abundance ratio directly using the \fexxv\ resonance line $w$ (transition $1s^2\,^1S_0 - 1s2p\,^1P_1$), which is frequently the strongest line making up the 6.7~keV line complex, and a feature at 7.06~keV (1.76~\AA) made up of Fe K$\beta$ lines, with inner-shell transition ($1s-3p$) and formed by fluorescence of the photosphere by the coronal hot flare plasma. Spectra from the Bragg Crystal Spectrometer on the Japanese {\em Yohkoh} spacecraft were used to find that the the flare Fe abundance was no more than a factor 2 higher than the photospheric abundance. A preliminary analysis \citep{phi95} of \fexxv\ and Fe K$\alpha$ line spectra observed by the {\sc solflex} instrument on {\em P78-1} spacecraft and the Bent Crystal Spectrometer (BCS) on {\em Solar Maximum Mission} ({\em SMM}) points to the photospheric and coronal Fe abundances being equal.

Here the analysis of \cite{phi95} is revisited, with the theory and observations of the ratio of the Fe K$\alpha$ lines to the \fexxv\ $w$ re-examined. The original observations from the SOLFLEX instrument on the {\em P78-1} spacecraft are used, together with newly analyzed observations from the {\em Solar Maximum Mission} Bent Crystal Spectrometer. The photospheric-to-flare abundance ratios are now consistent with the flare Fe abundance estimates from {\em RHESSI}.

% S 2: Theory of line formation
\section{Theory of line formation}

The inner-shell K$\alpha$ and K$\beta$ lines visible in solar flare X-ray spectra are formed when K ($nl=1s$) shell electrons in the neutral or once-ionized Fe atoms in the photosphere are removed. The K-shell electron may be ejected by either electrons or X-ray photons having energies $> 7.11$~keV. The vacancy in the K-shell is filled by an electron in the L ($nl=2p$) or M ($nl=3p$) shell; this results in either the emission of a photon or a re-arrangement of the remaining electrons with the emission of an Auger electron. In the case of the L--K transition, the photon emission forms two lines, with K$\alpha_1$ at energy 6.404~keV (1.936~\AA, transition $1s_{1/2} - 2p_{3/2}$) and K$\alpha_2$ at energy 6.391~keV (1.940~\AA, $1s_{1/2} - 2p_{1/2}$). Their intensity ratio is K$\alpha_1$:K$\alpha_2 = 2:1$. The M--K transition results in K$\beta$ line emission, with the most intense components being the K$\beta_1$ and K$\beta_3$ lines at 7.057~keV (1.76~\AA). The relative probability of photon emission is expressed by the fluorescence yield $\omega_K$ which for Fe is 0.31 \citep{fin66}. This rapidly increases with atomic number (roughly as $Z^4$), so explaining the fact that the fluorescence lines of Fe are the only ones ever to be observed with solar X-ray crystal spectrometers (Ni, with higher $Z$ and larger $\omega_K$, has an abundance about 20 times less than Fe).

For solar flares, the Fe inner-shell lines can be excited by energetic electrons \citep{phi73}, such as those in the tail of the Maxwell-Boltzmann distribution in hot flare plasmas or nonthermal electrons present at the flare impulsive stage, or by photons from the flare plasma incident on the photosphere \citep{bai79}. In the case of the latter, a strong dependence on the flare heliocentric angle would be expected, with limb flares having zero or very small fluorescence emission. This centre-to-limb dependence has been observed from spectral data from the {\em SMM} and {\em Yohkoh} crystal spectrometers \citep{par84,phi94}, and so it may be presumed that fluorescence emission predominates. Indeed, only one ambiguous case has been cited of a flare during its impulsive stage for which nonthermal electron excitation of the Fe K$\alpha$ lines has been observed \citep{ems86}.

The Fe K$\alpha$ or K$\beta$ fluorescence emission is formed in a layer of the solar atmosphere where X-rays with energies $> 7.11$~keV are absorbed by Fe atoms or ions. The absorption cross sections above and below the 7.11~keV edge by Fe indicate that absorption effectively occurs at a mass column density of $\sim 1$~g~cm$^{-2}$, which in the VAL \citep{ver81} ``average" model solar atmosphere and for the photospheric Fe abundance \citep{asp09,caf11} is located at about 200~km above the $\tau_{5000} = 1$ level, on the photospheric side of the temperature minimum.

The theory for the formation of the Fe K lines by fluorescence has been described by \cite{bai79}, whose theory was based on a Monte Carlo analysis. We briefly outline it here. The flux of the K$\alpha$ lines at the distance of the Earth $D$ is

% Eq. 1
\begin{equation}
F({\rm K}\alpha) = \frac{B \,\Gamma f(\theta, h)}{1 + \alpha f(\theta, h)} \frac{1}{4 \pi D^2} \int_{7.11}^{\infty} I_{\rm >7.11}\, EM (E) dE  \,\,\,\,\,{\rm photons\,\,\,cm}^{-2}\,\,{\rm s}^{-1}
\end{equation}

\noindent where the branching ratio $B$ for K$\alpha$ line emission is 0.882 (0.118 for K$\beta$ line emission: \cite{bam72}), $\Gamma$ is the efficiency of the production of K$\alpha$ line emission (equal to the ratio of line emission to the total emission at energies greater than the K-shell ionization energy, 7.11~keV), and $f(\theta, h)$ is a function describing the dependence of the line emission on the heliocentric angle $\theta$ and height $h$ above the photosphere of the hot flare plasma fluorescing the photosphere. The factor $[1+\alpha f(\theta, h)]$ corrects for Compton-scattered albedo photons contributing to the total emission with energy $> 7.11$~keV, $I_{\rm >7.11}$ times the flare's volume emission measure $EM$. This emission is taken to be thermal in \cite{bai79}'s analysis, and is made up of free--free and free--bound continua, with negligible line emission. In this analysis, $I_{\rm >7.11}$ was taken to be the continuum emission given by the {\sc chianti} database and software package \citep{der97,der09} rather than the analytical expression derived by \cite{phi94}. The emission measure $EM$ is $N_e^2 V$ where $N_e$ is the electron density and $V$ the flare's emitting volume. The fluorescence efficiency $\Gamma$ is a function of the flare plasma height $h$, the photospheric Fe abundance $N({\rm Fe})/N({\rm H})$, and the temperature $T_e$ of the hot plasma, assumed isothermal. The photospheric Fe abundance affects the value of $\Gamma$ in a nearly linear way; $A({\rm Fe}) = 7.60$ (or $N({\rm Fe})/N({\rm H}) = 4.0 \times 10^{-5}$) was used in \cite{bai79}'s analysis. We adapted this using $\Gamma ' = \Gamma \times [N({\rm Fe})/N({\rm H})]_{\rm phot} / [N({\rm Fe})/N({\rm H})]_{\rm Bai}$ where $[N({\rm Fe})/N({\rm H})]_{\rm Bai} = 4.0 \times 10^{-5}$ (corresponding to $A({\rm Fe}) = 7.60$) and $[N({\rm Fe})/N({\rm H})]_{\rm phot}$ was set equal to $3.2 \times 10^{-5}$, corresponding to \cite{asp09}'s value ($A({\rm Fe}) = 7.50$).

The \fexxv\ $w$ line is emitted by the flare plasma by collisional excitation processes. The $w$ line flux is then given by

% Eq. 2
\begin{equation}
F(w) = \frac {G(T_e) \times EM'}{4 \pi D^2}
\end{equation}

\noindent with $G(T_e)$ the contribution function for the $w$ line including high-$n$ dielectronic satellite lines that are spectroscopically indistinguishable from line $w$. The emission measure $EM'$ is taken to be equal to $EM$ for the emission at energies $> 7.11$~keV, i.e. the emitting plasma is assumed to be isothermal. The justification for this is from {\em RHESSI} flare spectra which show this to be a good approximation for flares in their decay stage \citep{phi11}. For the \fexxv\ $w$ line alone, $G(T_e)$ is given by

% Eq. 3
\begin{equation}
G(T_e) =  \frac{N({\rm Fe}^{+24}_e)}{N({\rm Fe}^{+24})} \frac{N({\rm Fe}^{+24})}{N({\rm Fe})} \Big[\frac{N({\rm Fe})}{N({\rm H})}\Big]_{fl} \frac{N({\rm H})}{N_e}  \frac{A_{i0}}{N_e} \,\,\,\,\,\,{\rm cm}^3\,\,{\rm s}^{-1}
\end{equation}

\noindent where $N({\rm Fe}^{+24}_e)$ is the  population of the excited level of the He-like Fe ion, $N({\rm Fe}^{+24})/{N({\rm Fe})}$ is the He-like ion fraction.  $[N({\rm Fe})/N({\rm H})]_{fl}$ is the coronal flare Fe abundance relative to H, and $A_{e0}$ the transition probability from level $e$ to the ground state; $N({\rm H})/N_e$ is taken to be 0.83. $G(T_e)$ was evaluated from {\sc chianti}, with ion fractions from \cite{bry09}, and the flare Fe abundance  $[N({\rm Fe})/N({\rm H})]_{fl}$ (see below). Although {\sc chianti} has data for unresolved satellites converging on to the \fexxv\ $w$ line, these include satellites of the type $1s^2 nl - 1s 2p nl$ with $nl$ with principal quantum number $n$ up to 5. We instead used data from \cite{bel82} which includes satellites with much higher $n$ values; this is a decreasing function of $T_e$, e.g. equal to 50\% of the \fexxv\ $w$ line $G(T_e)$ for $T_e = 17$~MK. We denote the sum of the $G(T_e)$ function for the \fexxv\ $w$ line and unresolved \fexxiv\ satellites by $G'(T_e)$.

The ratio of fluxes in the Fe K$\alpha$ lines to the \fexxv\ $w$ line plus unresolved \fexxiv\ satellites is then

% Eq. 4
\begin{equation}
\frac{F({\rm K}\alpha)}{F(w)} = R = \frac{B\, \Gamma ' f(\theta, h)}{1 + \alpha f(\theta, h)} \frac{\int_{7.11}^{\infty} EM \,\, I_{\rm >7.11} (E) dE}{G'(T_e)}
\end{equation}

\noindent The ratio $R$ is a function of $T_e$, the flare heliocentric angle $\theta$ and its height above the photosphere $h$. It is also proportional to the ratio of the photospheric Fe abundance $[N({\rm Fe})/N({\rm H})]_{\rm phot}$ (through $\Gamma '$) to the flare plasma Fe abundance, $N({\rm Fe})/N({\rm H})_{fl}$ through the $G(T_e)$ function (Eq.~3), neglecting the contribution of unresolved satellites.

\begin{figure}
\begin{center}
\epsfysize=6.5cm
\epsfbox{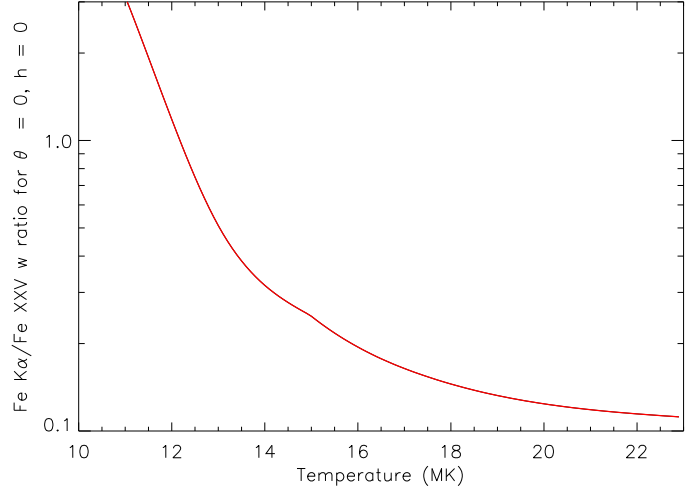}
\caption{Dependence of the ratio $R$ of Fe K$\alpha$ line flux to \fexxv\ line $w$ line flux for disk-centre ($\theta = 0$) and zero height flares.  } \label{Kalpha_ratio_temp_dep}
\end{center}
\end{figure}

The ratio $R$ is plotted logarithmically as a function of $T_e$ for disk-centre ($\theta = 0$) and $h=0$ flares with photospheric and flare Fe abundances set equal (Figure~\ref{Kalpha_ratio_temp_dep}). There is a steep decline for temperatures that are relatively low but for those considered here, $T_e = 17$~MK, attained shortly after flare peak values, the dependence is more gradual.

% Figure 1: dependence on T of K-alpha/w line ratio for disk-centre flares.

\begin{figure}
\begin{center}
\epsfysize=6.5cm
\epsfbox{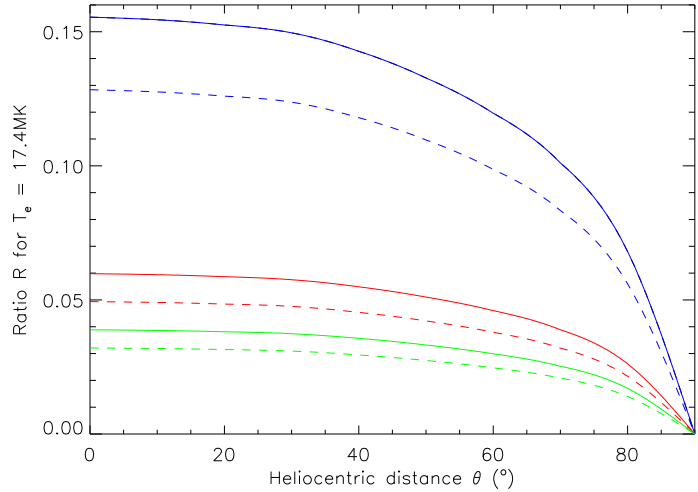}
\caption{Dependence of the ratio $R$ for $T_e = 17.4$~MK on heliocentric angle $\theta$ and flare abundances, equal to the abundances of Feldman \& Laming (2000) (blue), Phillips \& Dennis (2011) (red), and Asplund et al. (2009) (green). The dependence on flare heights is shown by the $h=0$ (solid curves) and $h=7000$~km (dashed curves). } \label{Kalpha_ratio_abund_theta}
\end{center}
\end{figure}

The dependence of $R$ on the flare Fe abundance and $\theta$ is shown in Figure~\ref{Kalpha_ratio_abund_theta} for heights $h$ of 0 and 7000~km. The flare abundances $[N({\rm Fe})/N({\rm H})]_{fl}$ chosen are the coronal value of \cite{fel00}, the photospheric value of \cite{asp09}, and the value from {\em RHESSI} observations \citep{phi11}.

% Figure 2: dependence on theta of K-alpha/w line ratio

% S 3: Observations
\section{Observations}

Spectra over the 6.4--6.7~keV ($1.85-1.94$~\AA) range, including the \fexxv\ $w$ and Fe K$\alpha$ lines, have been made by a number of spacecraft crystal spectrometers operating between 1979 and 1989. The flat crystal spectrometer SOLFLEX on {\em P78-1} scanned the range repeatedly back and forth and obtained several spectra during flares \citep{dos79}. Measurements of line fluxes during a small sample of flares were made for the analysis of \cite{phi95}. Unfortunately, an on-line archive of the data is not available, so for the analysis here these measurements are used again. These observations allow direct comparison of the \fexxv\ $w$ and Fe K$\alpha$ lines which are clearly resolved.

\begin{figure}
\begin{center}
\epsfysize=6.5cm
\epsfbox{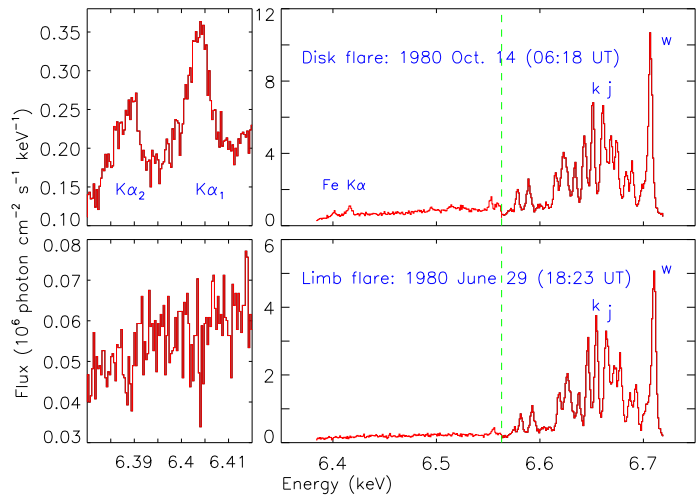}
\caption{{\em SMM} BCS channels~3 and 4 spectra (right panels) for a disk flare (1980 October 14, $\theta = 16^\circ$), showing the K$\alpha$ lines at 6.4~keV in channel~2 (left panel), and a limb flare (1980 June 29), with no visible K$\alpha$ line emission. The vertical dashed green lines indicate the boundaries of channels~3 and 4. The Fe~{\sc xxv} $w$ line (6.699~keV) and two of the Fe~{\sc xxiv} dielectronic satellites $j$ and $k$ are indicated. The region of the Fe K$\alpha$ lines as viewed by channel~2 are shown in the left panels. The spectra are plotted on an energy scale (abscissa) and absolute flux scale (ordinate) using the effective areas of Table~\ref{BCS_channels}.} \label{BCS_disk_limb_flare_sp}
\end{center}
\end{figure}

The Bent Crystal Spectrometer (BCS) on {\em SMM} consisted of 8 channels, 7 covering the Fe lines of interest. A bent germanium crystal (Ge 422 crystal plane) with particular bend radius for each of these channels allowed portions of the full range to be observed, the spectra being observed by position-sensitive proportional counters. The spectral ranges are given in Table~\ref{BCS_channels}, together with effective areas and instrumental line broadening parameters, all based on pre-launch measurements \citep{act80}. The main advantages of bent crystal spectrometers include the fact that, over a relatively short time (DGI or data gathering interval, typically a few seconds), photon counts making up a spectrum can be accumulated in each channel's range, whereas with a scanning flat crystal spectrometer particular spectral lines are scanned at slightly different times. This advantage is most apparent during a flare's rapid rise. Channels~4 and 7 of the BCS observed the \fexxv\ $w$ line, with channel~4 having a larger range but smaller sensitivity than channel~7. The Fe K$\alpha$ lines were observed in channels~2 and 3, with channel~3 having a larger wavelength range but smaller sensitivity than channel~2. Thus, to evaluate the flux ratio of the Fe K$\alpha$ and \fexxv\ lines involves the ratio of measured effective areas of two separate channels. For this analysis, individual line fluxes were measured by best-fit Voigt functions to the line profiles using routines written in Interactive Data Language (IDL), with the Gaussian part of the function equal to the thermal Doppler broadening plus electronic broadening and the Lorentzian part approximating the crystal rocking curve (this assumption is not critical as the rocking curve width is approximately 10 times less than the thermal Doppler broadening width). Figure~\ref{BCS_disk_limb_flare_sp} shows spectra in channels 3 and 4 during two large flares seen by {\em SMM}, one near disk centre ($\theta = 16^\circ$) and the other near the west limb ($\theta = 90^\circ$). The Fe K$\alpha$ lines are visible only in the disk flare spectrum. These spectra are composites of channels 3 and 4. The Fe K$\alpha$ lines are much more conspicuous in the more sensitive channel~2 (see left panels of the figure). Fluorescence of the germanium material of the crystals by solar X-rays during flares is observed by the position-sensitive detectors as a background over all wavelengths, present in the spectra of Figure~\ref{BCS_disk_limb_flare_sp} as well as the SOLFLEX spectra.

% Table 1: SMM BCS channels
\begin{table*}
\begin{center}
%\rotate
\caption{{\em SMM} Bent Crystal Spectrometer Channels}
\label{BCS_channels}

\begin{tabular}{llclc}\hline\hline
\\
Chan. & Lines & Energy range (wavelength range) & Effective area & Instrumental \\
No.& included & keV (\AA)& (cm$^2$) & broadening \\
& & & & (FWHM) in keV [\AA]$^1$ \\

2   & Fe K$\alpha$ & 6.375--6.431 (1.928--1.945) & 0.019 & 4.89 (-4) [1.48(-4)] \\
3   & \fexxv\ $w$ to \fexxi\ sats. & 6.368--6.550 (1.893--1.947) & .0063& 1.47 (-3) [4.36(-4)] \\
4   & \fexxi\ sats. to Fe K$\alpha$ & 6.546--6.739 (1.840--1.894) & .0059& 1.55 (-3) [4.35(-4)] \\
7   & \fexxv\ $w$ & 6.684--6.731 (1.842--1.855) & .025 & 6.64 (-4) [1.83(-4)] \\
\\ \hline
\end{tabular}
\end{center}
\footnote{a} Quadratic sum of crystal rocking curve and electronic broadening.
\end{table*}

% Fig. 3: SMM BCS spectra, limb and disk flares

For both the {\em P78-1} and {\em SMM} observations, large flares were selected with a variety of heliocentric distances. As both SOLFLEX and BCS were only coarsely collimated spectrometers, they do not give accurate solar coordinates for each flare;  these were instead obtained from the coordinates of H$\alpha$ flares listed in the {\em Solar-Geophysical Data Bulletin}. Generally the correlation of the X-ray and H$\alpha$ flares was unambiguous but at intervals of high solar activity there were occasionally more than one H$\alpha$ flare occurring simultaneously; these flares were rejected from the data base. Table~\ref{flare_list} lists the flares observed by SOLFLEX and BCS included in the analysis. Most are in 1980, near the peak of Cycle~22. Although {\em SMM} continued operating after 1984, when the spacecraft attitude control unit was replaced by Space Shuttle astronauts, activity had declined and BCS channel~2 had failed, so that there were few flares available for analysis.

% Table 2: List of flares with Fe K-alpha fluxes from SMM BCS
\begin{table*}
\begin{center}
%\rotate
\caption{Flares with Fe K$\alpha$ line emission}
\label{flare_list}

\begin{tabular}{llcrrrr}\hline\hline
\\
{\em SMM} BCS Flares \\
Flare no. & Date & Time (UT) & Ch. 7 $w$ & Ch. 2 K$\alpha$ & Fe K$\alpha$/\fexxv\ $w$ & Heliocentric \\
& & count rate$^1$ & count rate & count rate & ratio $R_{\rm obs}$ & distance $\theta$ ($^\circ$) \\

1...& 1980 Apr 7  & 05:40 &  604 (6) & 40 (2) & 0.087 (6) & 16 \\
2...& 1980 Apr 10 & 09:18 &  452 (7) & 20 (1) & 0.058 (4) & 46 \\
3...& 1980 May 9  & 07:13 & 1057 (10) & 68 (2) & 0.085 (4) & 36 \\
4...& 1980 May 21 & 21:01 & 1689 (14) & 77 (2) & 0.060 (2) & 17 \\
5...& 1980 Jun 25 & 15:54 &  512 (8) & 27 (1) & 0.069 (4) & 42 \\
6...& 1980 Jul 1 & 16:31  &  190 (4) & 6 (1) & 0.038 (10) & 40 \\
7...& 1980 Jul 5 & 22:49 &  857 (10) & 66 (3) & 0.101 (9) & 37 \\
8...& 1980 Jul 12 & 11:17 & 257 (5) & 14 (1) & 0.069 (7) & 58 \\
9...& 1980 Jul 14 & 08:28 & 494 (7) & 33 (2) & 0.088 (6) & 45 \\
10...& 1980 Jul 21 & 03:00 & 491 (7) & 23 (1) & 0.062 (4) & 63 \\
11...& 1980 Aug 23 & 21:28 &  53 (3) & 3 (1) & 0.065 (25) & 49 \\
12...& 1980 Aug 31 & 12:51 & 118 (4) & 9 (1) & 0.100 (16) & 29 \\
13...& 1980 Sep 24 & 07:35 & 172 (4) & 9 (1) & 0.065 (10) & 27 \\
14...& 1980 Oct 14 & 06:19 & 1719 (14) & 87 (3) & 0.067 (3) & 16 \\
15...& 1980 Nov 5 & 22:35 &  525 (4) & 33 (1) & 0.083 (4) & 10 \\
16...& 1980 Nov 7 & 04:55 &  182 (2) & 11 (1) & 0.076 (8) & 58 \\
17...& 1980 Nov 11 & 21:00 & 287 (3) & 18 (1) & 0.083 (6) & 16 \\
18...& 1980 Nov 12 & 02:52 & 139 (2) & 13 (1) & 0.123 (14) & 17 \\
19...& 1980 Nov 12 & 17:05 &  45 (1) & 2 (1) & 0.051 (11) & 20 \\
\\

{\em P78-1} SOLFLEX Flares\\
Flare no. & Date & Time (UT) & Fe K$\alpha$/\fexxv\ $w$ & Heliocentric \\
          &      &           & ratio $R_{\rm obs}$      & distance $\theta$ ($^\circ$) \\

1...      & 1979 Mar 22 & 13:43 & 0.057 & 30 \\
2...      & 1979 Mar 25 & 18:05 & 0.043 & 79 \\
3...      & 1979 May  2 & 17:03 & 0.049 & 60 \\
4...      & 1979 Nov  8 & 23:35 & 0.065 & 49 \\
5...      & 1980 May 18 & 01:11 & 0.061 & 26 \\
6...      & 1981 Apr 18 & 18:06 & 0.066 & 26 \\
\\

\hline

\end{tabular}
\end{center}
\footnote{1} {Numbers in parentheses are uncertainties in the last significant figure. }
\end{table*}

% Section 5
\section{Results}

Measured flux ratios $R_{\rm obs}$ of the Fe K$\alpha$ to the \fexxv\ $w$ line for times when the estimated temperature from the flux ratio of the \fexxiv\ $j$ satellite to the \fexxv\ $w$ line was 17~MK (corresponding to a $j/w$ ratio of 0.58) are given in Table~\ref{flare_list}. A typical DGI for the observations was 16~s. For the {\em SMM} BCS measurements, fluxes in channels~2 and 7 only are given. The uncertainties given in the table are statistical only, estimated from the Voigt line fitting routine; they do not include other possible uncertainties in the fitting process, e.g. in placing the fluorescence background. The uncertainties in the derived ratios result from propagating those in the measured Fe K$\alpha_1$, Fe K$\alpha_2$, and \fexxv\ $w$ line fluxes. Corresponding measurements from channel~3 have much larger uncertainties as a result of the smaller photon count rates in this channel, as can be seen from Figure~\ref{BCS_disk_limb_flare_sp}, and are not given in the Table. For the {\em P78-1} SOLFLEX observations, a notional uncertainty of 20\% is assigned, based on the estimation of statistical noise of the spectra from our earlier analysis \citep{phi95}.

\begin{figure}
\begin{center}
\epsfysize=6.5cm
\epsfbox{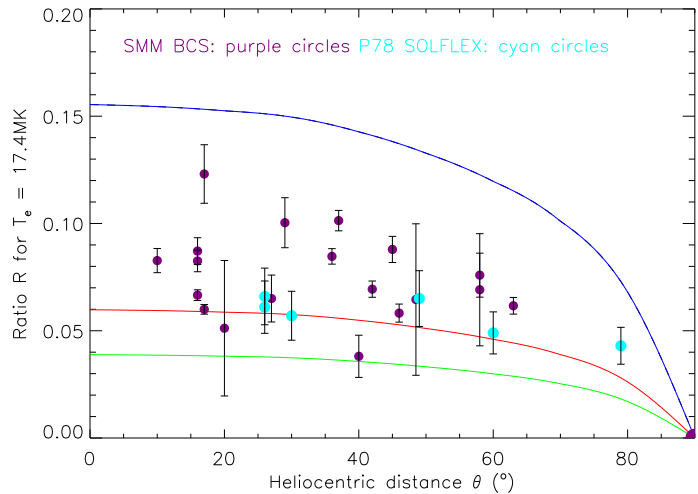}
\caption{Observed ratios of Fe K$\alpha$ line flux ({\em SMM} BCS channel 2) to \fexxv\ $w$ line flux (channel~7) during flares listed in Table~2 plotted against flare heliocentric angle $\theta$ compared with theoretical curves calculated for zero height and for photospheric Fe abundance (blue), the flare abundance of Phillips \& Dennis (2011) (red), and the coronal Fe abundance of Feldman \& Laming (2000) (green). The {\em SMM} BCS points are in brown, the {\em P78-1} SOLFLEX points are in cyan (light blue). The corresponding $h=7000$~km curves (not shown) are about 20\% lower than the solid curves in this plot.} \label{Kalpha_ratio_ch27}
\end{center}
\end{figure}

The ratios are plotted against the flare heliocentric distance in Figure~\ref{Kalpha_ratio_ch27} with zero-height theory curves for the three Fe abundances chosen from Figure~\ref{Kalpha_ratio_abund_theta}. The justification for choosing zero height is from {\em Skylab} images with the S082A slitless extreme ultraviolet spectrograph (e.g. \cite{wid74}) which imaged a number of flares in the 192~\AA\ \fexxiv\ line, showing very small heights at the temperature of this line which is almost as hot as the \fexxv\ $w$ line. A number of limb flares were observed with no evidence for the Fe K$\alpha$ lines; they are shown as the single point with zero ratio at $\theta = 90^\circ$. There is near-agreement of the six {\em P78-1} points with the Fe abundance of \cite{phi11}, determined from {\em RHESSI}. For the 19 {\em SMM} points, there is larger scatter, with the points spread between the curves for the Fe abundance of \cite{phi11} and that of \cite{fel00}. We note that the theoretical curves in Figure~\ref{Kalpha_ratio_abund_theta} for $h=7000$~km would be about 20\% lower than the appropriate solid curves, so if the mean flare height is of this order, the coronal Fe abundance is slightly less.

\begin{figure}
\begin{center}
\epsfysize=6.5cm
\epsfbox{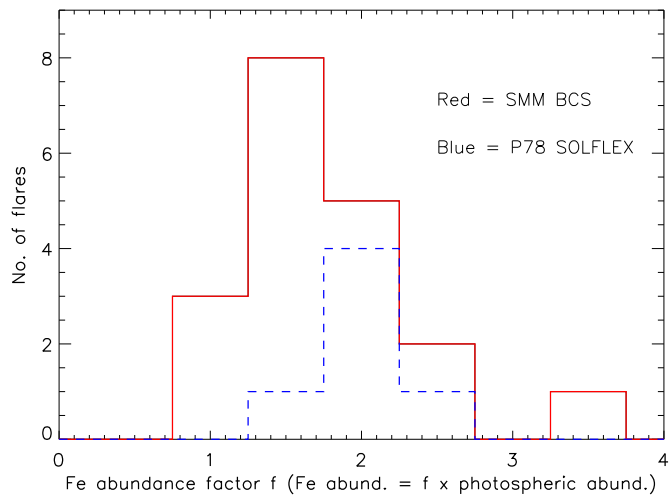}
\caption{Histograms of Fe abundance factors $f$ from {\em SMM} BCS and {\em P78-1} SOLFLEX spectra. The Fe abundance from each observation is $f$ times the photospheric abundance of Asplund et al. (2009).  } \label{histogram}
\end{center}
\end{figure}

An estimate of the Fe abundance can be made from the {\em SMM} and {\em P78-1} ratios by comparing for each observation the factor $f$ that the photospheric Fe abundance must be multiplied by to obtain the measured ratio. For the {\em SMM} points, $f$ varies from 1.25 to 3.75, and for the {\em P78-1} points from 1.70 to 2.62. Histograms of the two distributions of $f$ are given in Figure~\ref{histogram}. Best-fit Gaussian curves to each distribution indicates that $f$ for the {\em SMM} points is $1.6\pm 0.5$ and for the {\em P78-1} points $2.0 \pm 0.3$. There is thus some evidence for a small ($\sim 24$\%) systematic difference between the {\em SMM} and {\em P78-1} ratios.

% Fig. 4: Plot of Fe K-alpha to Fe XXV w line for SMM BCS & P78 SOLFLEX spectra

% Fig. 5: Histogram of Fe abundance factors f

\begin{figure}
\begin{center}
\epsfysize=6.5cm
\epsfbox{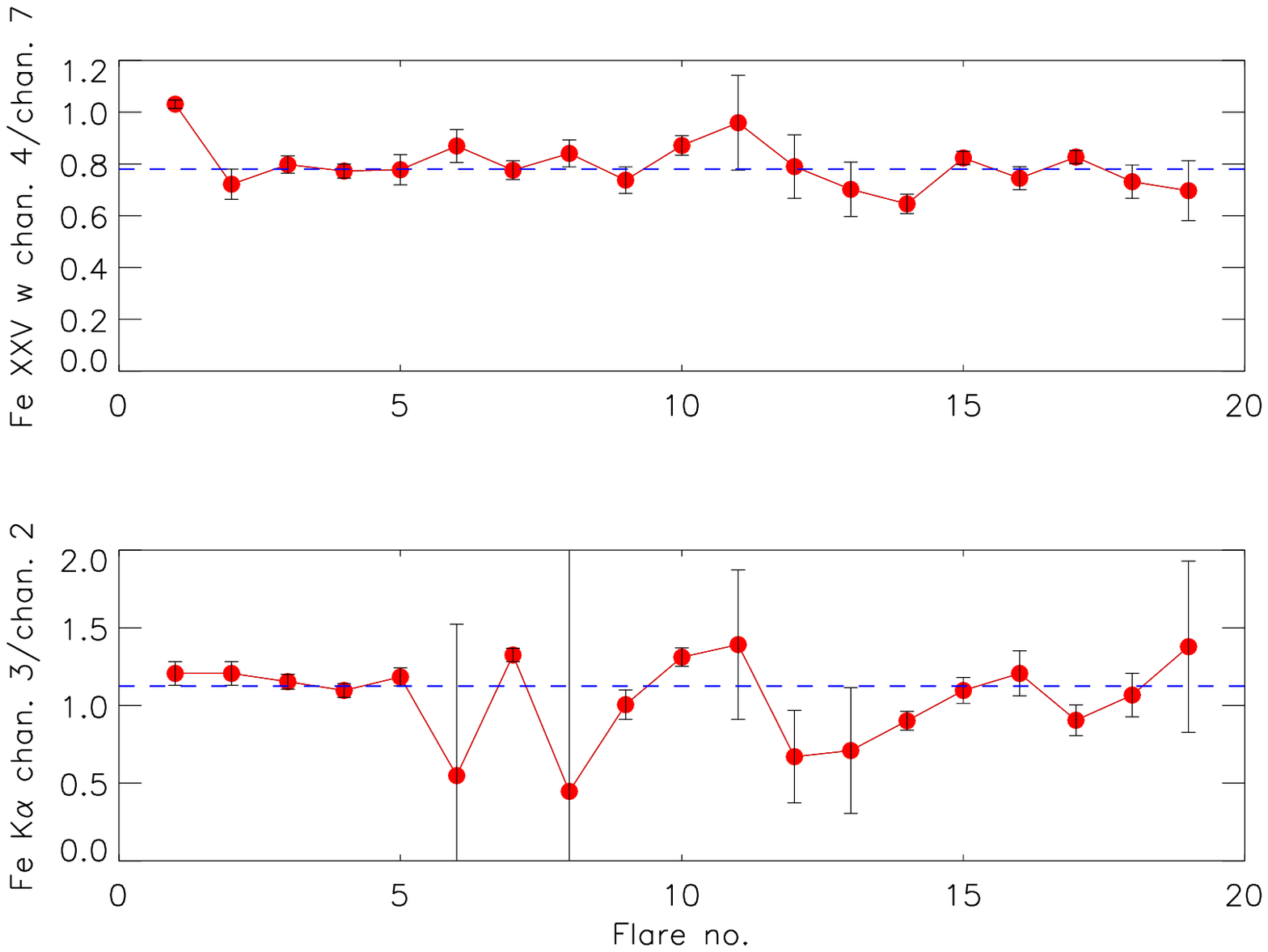}
\caption{Upper panel: Ratio of fluxes in the \fexxv\ line $w$ estimated from {\em SMM} BCS channel~4 to channel~7 photon count rates with effective areas from Table~\ref{BCS_channels}. The mean ratio for these 19 observations is $0.78 \pm 0.09$ (blue horizontal line). Lower panel: Ratio of fluxes in the Fe K$\alpha$ lines estimated from {\em SMM} BCS channel~3 to channel~2 photon count rates and effective areas. The line is weak in particularly channel~3 and so the mean ratio, $1.12 \pm 0.35$ (blue horizontal line), is poorly determined. In each case, fluxes were determined using effective areas in Table~\ref{BCS_channels}. } \label{BCS_channel_ratio}
\end{center}
\end{figure}

A source of this systematic difference is very likely to be the effective areas of the {\em SMM} BCS channels 2 and 7. In the analysis of \cite{phi95}, a rough estimate of the relative effective area uncertainties was given as $\sim 30$\%. Some corroboration of this is provided here by simultaneous measurements of the \fexxv\ $w$ line in channels~4 and 7, the line being strong in both channels despite the factor-of-4 smaller sensitivity of channel 4. The channel~4 to channel~7 ratio is plotted against flare number in Figure~\ref{BCS_channel_ratio} (upper panel). Evidently the ratio is at least 20\% less than one as expected if the measured effective areas were precisely correct. The channel~3 and channel~2 measured fluxes of the K$\alpha$ lines are more difficult to compare because of the weakness of the lines in channel~3 (lower panel). The ratio is consistent with unity (mean value 1.12), but the statistical uncertainty is $\sim 35$\%. In summary, the ratio of relative sensitivities of BCS channels 2 and 7 probably accounts for the discrepancy between the {\em SMM} and {\em P78-1} points in Figure~\ref{Kalpha_ratio_ch27}.

% Fig. 6: Plot of BCS Fe XXV w line ch. 4/ch. 7 ratio and Fe K-alpha ch. 3/ch. 2 ratios

% Section 6
\section{Conclusions}

The possibility of obtaining the flare-to-photospheric Fe abundance ratio using the fluorescence-formed Fe K$\alpha$ and K$\beta$ lines and the \fexxv\ $w$ line, expressed in our earlier work \citep{phi94,phi95}, is carried further here in the case of {\em SMM} Bent Crystal Spectrometer and {\em P78-1} SOLFLEX spectra in the 6.39--6.70~keV (1.85--1.94~\AA) range. The six {\em P78-1} measurements of the Fe K$\alpha$-to-\fexxv\ $w$ line ratio indicate a flare-to-photospheric Fe ratio of $2.0 \pm 0.3$, while the 19 {\em SMM} measurements indicate a flare-to-photospheric Fe abundance ratio of $1.6 \pm 0.5$. The slight difference is likely to be due to $\gtrsim 20$\% errors in the relative effective areas of BCS channels~2 and 7. The flare-to-photospheric Fe ratio from the {\em P78-1} measurements alone is consistent, to within uncertainties, with recent estimates of the flare Fe abundance from {\em RHESSI} spectra \citep{phi11} which are a factor $2.6 \pm 0.6$ higher than the photospheric Fe abundance \citep{asp09,caf11}. This ratio is also consistent with {\em Yohkoh} Bragg Crystal Spectrometer measurements of the Fe K$\beta$ line \citep{phi95}. These measurements of the Fe abundance enhancement may be relevant to the recent result of \cite{woo10} suggesting that FIP enhancements in active stars are greatest for Sun-like stars (spectral type G), decrease to zero for cool stars of type K0, while for later spectral types the coronal abundances of low-FIP elements are diminished relative to photospheric. The zero enhancement of Fe in the corona above a sunspot \citep{fel90} with effective temperature approximately equal to a K0 star and our own result for solar flares may fit such a picture.

\section*{Acknowledgments}

Much of the analysis of {\em SMM} BCS data has relied on the IDL routines written by D. M. Zarro, to whom thanks are due. I thank U. Feldman for use of the {\em P78-1} SOLFLEX data. {\sc chianti} is a collaborative project involving the US Naval Research Laboratory, the Universities of Florence (Italy) and Cambridge (UK), and George Mason University (USA).

\bsp

\label{lastpage}

\end{document}